\documentclass[11pt,twocolumn]{article}
\usepackage{cite}
\usepackage{graphicx}
\usepackage{pdfpages}
\usepackage{rotating}
\usepackage{placeins}
\usepackage{csvsimple}
\usepackage{listings}
\usepackage{url}
\usepackage{amsmath}
\usepackage{stmaryrd}

\usepackage{siunitx}
\usepackage{booktabs}
\usepackage{xcolor}
\usepackage{setspace} 
\usepackage{verbatim}
\usepackage{textcomp}

\providecommand{\keywords}[1]{\textbf{\textit{Index terms---}} #1}

\begin{document}

%\doublespacing

\title{The INSuRE Project:\\
CAE-Rs Collaborate to Engage Students in Cybersecurity Research}

%% Areas to redact:
%% authors, acks, names in text (Dark, footnote), UMBC, pubs
%% school lists

%\author{Redacted for anonymous submission}

%\begin{comment}
\author{Alan Sherman (UMBC),
M. Dark (Purdue),
A. Chan (Northeastern),\\
R. Chong (Purdue),
T. Morris (UAH), 
L. Oliva (UMBC),
J. Springer (Purdue),\\
B. Thuraisingham (UTD),
C. Vatcher (UMBC),
R. Verma (Houston),
S. Wetzel (Stevens)$^{1}$
}
%\end{comment}

\date{\today}

\maketitle

\footnotetext[1]{University of Maryland, Baltimore County (UMBC),
University of Alabama in Huntsville (UAH),
University of Texas at Dallas (UTD),
University of Houston (Houston),
Stevens Institute of Technology (Stevens)}
\setcounter{footnote}{1}
% Purdue University (Purdue)
% Northeastern University (Northeastern)

\begin{abstract} % 62 words

Since fall 2012, several 
National Centers of Academic Excellence in Cyber Defense Research (CAE-Rs)
fielded a collaborative course
to engage students in solving applied cybersecurity research problems.  
We describe our experiences with this
Information Security Research and Education (INSuRE) research collaborative.
We explain how we conducted our project-based research course,
give examples of student projects, and
discuss the outcomes and lessons learned.

\end{abstract}


\section{The INSuRE Project}
\label{sec:intro}

The {\it Information Security Research and Education (INSuRE)} research 
collaborative\footnote{\url{http://insurehub.org/about-us }} 
is a network of 
{\it National Centers of Academic Excellence in Cyber Defense Research (CAE-Rs)} universities
that cooperate to engage students in solving applied cybersecurity research problems.  
Since fall 2012, INSuRE has fielded a multi-institutional cybersecurity research course in which
BS, MS, and PhD students work in small groups to solve unclassified problems proposed by 
the {\it National Security Agency (NSA)} and by other government and private organizations and laboratories.

In this paper we describe our experiences with the INSuRE Project.
We explain how we conducted our project-based research course,
give examples of student projects, and
discuss the outcomes, benefits, and lessons learned.

The approximately eighty CAE-R universities\footnote{\url{https://www.iad.gov/NIETP/reports/cae_designated_institutions.cfm }}
include a significant collection of cybersecurity students, educators, and researchers.
While the individual universities were ``nodes of excellence,'' these nodes were not
purposefully constellated into a research network.
The INSuRE Project created an educational and research network of CAE-Rs.
As such, INSuRE is a self-organizing, multi-disciplinary, multi-institutional, and multi-level collaborative organization. 

The central activity of the INSuRE Project is its cybersecurity research course, in which students form small groups that work on
research problems of interest to the nation.  The NSA and other organizations support the project by contributing 
suggested problems and by providing technical directors to mentor student groups.  The geographically-diverse
participants connect and collaborate using a variety of conferencing and data-sharing technologies.

Students benefit from an exciting opportunity to work collaboratively on real-world problems 
and to interact with experienced technical directors.
They learn how to carry out research, including producing fast, incremental, and actionable results in team projects.  
Benefits to participating government organizations include collaborative
work on important problems and access to university faculty and to highly motivated and capable students
for possible employment.
In addition, faculty benefit from
building connections with other researchers, schools, and government organizations.

Melissa Dark (Purdue), and Mark Loepker, NSA Security Education Academic Liaison (SEAL) for Purdue, 
started the INSuRE Project in 2012.   
Dark provided the students and Loepker provided the NSA research problems, 
along with NSA Technical Directors under Trent Pitsenbarger's leadership.  
For more information about INSuRE, see Dark, et al.~\cite{dark2014,dark2015a,dark2015b}.

%insure2017
% redacted Dark in citation

\section{The INSuRE Course}
\label{sec:course}

%%%%%%%%%%%%%%%%%%%%%%%%%%%%%%%%%%%%%%%%%%%%%%%%%%%%%%%%%%%%%%%%%%
\begin{table}[!t]

\caption{Growth of the INSuRE cybersecurity research course
from fall 2012 through spring 2017:
number of universities, 
students,  technical directors (TDs: NSA + other government organizations), 
and student groups.
For example, in fall 2016, there were eleven TDs, including
five from NSA and six others.}

\label{tbl:history}

\begin{center}
\begin{tabular}{l|r|r|r|r}
	Term  	 		& Univs. 	& Stud- 	& TDs 		& Groups \\
					& 		 	& ents 		&  		&  \\
	\hline 
	2012 fall		& 1 		& 5 		& 3$+$0		& 1 \\
	2013 summer		& 1 		& 1 		& 0 		& 1 \\
	2014 spring		& 3 		& 33 		& 7$+$0		& 13 \\
	2014 summer		& 1 		& 1 		& 0 		& 1 \\
	2014 fall		& 4 		& 22 		& 7$+$1 	& 8 \\
	2015 spring		& 8 		& 52 		& 7$+$3 	& 21 \\
	2015 fall		& 7 		& 42 		& 8$+$8		& 14 \\
	2016 spring		& 6 		& 72 		& 6$+$9 	& 27 \\
	2016 fall		& 8 		& 64 		& 5$+$6 	& 25 \\
	2017 spring		& 7			& 54		& 6$+$15	& 22 \\
\end{tabular}

\end{center}
\end{table}
%%%%%%%%%%%%%%%%%%%%%%%%%%%%%%%%%%%%%%%%%%%%%%%%%%%%%%%%%%%%%%%%%%

% redact Dark

The first INSuRE course took place in fall 2012 at Purdue, involving five students who
formed two groups supported by three technical directors (TDs).  
With funding from the National Science Foundation,
the project added three more schools: 
University of California, Davis; Mississippi State; and University of Maryland, Baltimore County (UMBC).
Many of the INSuRE students are 
CyberCorps{\small\textregistered}: Scholarship for Service (SFS)\footnote{\url{https://www.sfs.opm.gov/ }} scholars.

In the following years, the course expanded to include a total of twelve universities,
six national labs, and two state organizations.\footnote{The additional 
universities included
Carnegie Mellon University,
Dakota State University,
Iowa State University,
Northeastern University,
Stevens Institute of Technology,
University of Alabama in Huntsville,
University of Houston,
University of Texas at Dallas, and 
University of Texas at San Antonio.
The additional organizations included
Argonne National Laboratories,
Indiana Office of Technology,
Johns Hopkins University Applied Physics Laboratory,
National Institute of Standards and Technology,
Naval Surface Warfare Center Crane Division,
New Jersey Office of Homeland Security and Preparedness,
Oak Ridge National Laboratories,
Pacific Northwest National Laboratories, and
Sandia National Laboratories.
}
In some years, a small number of private companies participated.  For example, the spring 2014 edition
included a private defense contractor, Assured Information Security, located in UMBC's research park.
Each partner organization suggested research problems.
Table~\ref{tbl:history} summarizes the growth of the INSuRE course.

Every semester, a rotating subset of the collaborating universities offers a section of the course at their 
schools.  Doing so enables each university to participate at a frequency that suits its needs, while 
fostering a diverse set of relationships among the schools.

To facilitate collaboration, the project used PURR, an instantiation at Purdue of the open source software platform 
HUBzero{\small\textregistered}.\footnote{\url{https://hubzero.org/ }}  
Users can share files, publish datasets and computational tools with Digital Object Identifiers, 
and participate asynchronously in discussion groups across multiple institutions.  
Individuals and groups participated synchronously in periodic community meetings using
the WebEx conferencing software, supplemented by an audio bridge.
INSuRE instructors shared experiences and developed common syllabi, handouts, and grading rubrics.

All class activities revolved around student projects.  TDs presented their suggested problems.
Students submitted bids and formed groups (typically three to five students each).  
In some schools, instructors assigned groups;  in other schools, students self-selected the groups.
Each group prepared a proposal, including a literature review, specific aims, and research plan.
Formal group presentations to the
INSuRE community included progress reports and final reports.  

Throughout the course, students interacted with their TD.   
TDs could check on the status of their groups, including
reading a ``dashboard'' slide summarizing the group's progress.

Most groups worked on problems suggested by the TDs;  some proposed their own ``custom'' projects
or variations of suggested problems.
Organizations sometimes proposed the same or a similar project in multiple semesters.
Student groups were allowed to continue projects completed in previous terms, or in some cases, 
revisit a problem addressed by others before.

Once a semester, key faculty and student members from each of the participating schools met in person, 
together with some of the TDs, to review outcomes,
discuss possible improvements, nurture relationships, and plan ahead.

In summer 2016, the project
initiated INSuRE-Con,\footnote{\url{https://sites.google.com/a/uah.edu/insurecon16/proceedings }}
an annual student-organized research conference featuring
five competitively-selected project presentations from the INSuRE class.

\section{Project Examples}
\label{sec:projects}

To illustrate the type of research carried out in the INSuRE course, 
we comment briefly on the suggested problem lists
and describe several representative projects,
including three in some detail.

\medskip \noindent {\it Problem Lists. }
Partner organizations provided
lists of suggested problems covering a wide range of topics, including, for example, 
policy-based stored information management, protection, and access control;
software assurance, including machine-assisted semantic understanding of code;
cloud computing, including cleaning up data spillage in Hadoop clouds;
forensics, including cloud forensics and mobility forensics in the Internet of Things;
deriving intelligence from an encrypted VPN stream;
protocol analysis and verification;
attacking botnets;
machine learning for malware classification;
vehicular data bus security; and 
incident response capabilities assessment.

TD Pitsenbarger explained,
``The tasks we place on our INSuRE task list represent areas 
where the organization needs greater insight and past tasks have helped us.''
These areas include 
understanding new technologies (e.g., FIDO authentication),
tool development (e.g., control flow integrity), and
validation of guidance (e.g., guidance on cleaning up spillages of sensitive data in clouds).

%%%%%%%%%%%%%%%%%%%%%%%%%%%%%%%%%%%%%%%%%%%%%%%%%%%%%%%%%%%%%%%%%%%%%%%%%%%%%%%%%
\medskip \noindent {\it Example 1: Moving Target Defense.}
% Bhavanai

A three-student team from UT Dallas, working together with Argonne National Laboratories, developed
a {\it Moving Target Defense (MTD)} to protect against probing attacks on web servers~\cite{tho2016}.
At random bounded intervals between 15 and 60 seconds, the system switched the web server
between Apache and Nginx.  Dynamically updated IP tables redirected web traffic to the active
server.  This deception aimed to hinder attacks by constantly changing the target.

To test the effectiveness of their system, the team launched simulated attacks against
the web service, with and without MTD protection.  With MTD protection, a Word Press application
ran normally 76\% of the time, 
experienced lag 14\% of the time, and
was down 10\% of the time.
By contrast, without MTD the application
ran normally 13\% of the time, 
experienced lag 7\% of the time, and
was down 80\% of the time.

This work suggests that MTD can be a practical and effective defense against
web service probing attacks.

%%%%%%%%%%%%%%%%%%%%%%%%%%%%%%%%%%%%%%%%%%%%%%%%%%%%%%%%%%%%%%%%%%%%%%%%%%%%%%%%%
\medskip \noindent {\it Example 2: Analysis of FIDO.}
% Sussane and Rylan

In three separate terms, 
teams from UMBC, Purdue, and Stevens Institute of Technology
analyzed the {\it Fast Identity Online (FIDO)} authentication protocol under
development by the FIDO Alliance.  In spring 2014, a team from UMBC 
studied the new FIDO protocol, assessing its goals, strengths, and weaknesses.
This team complemented the work of another concurrent UMBC team that studied
the PICO authentication system.  FIDO and PICO offer different approaches
toward the eventual replacement of passwords.  

%PICO requires a disruptive radical immediate paradigm shift to physical tokens;  
%FIDO offers a non-disruptive standard for gradually moving away from passwords.

Building on the initial UMBC work, in fall 2014, 
a team from Purdue evaluated the vulnerability of the
FIDO Ready (TM) Samsung Galaxy S5 fingerprint reader 
to a particular spoofing attack~\cite{cho2015}.
Discovered by a German security research lab, this attack
lifted a latent fingerprint. 
The Purdue team was unable to replicate this attack successfully,
though they were able to produce a fingerprint by
lifting a latent fingerprint.

Building on the Purdue work, in spring 2015,
a team from Stevens studied two attacks on 
each of four different fingerprint scanners: 
two FIDO compliant devices (Samsung Galaxy S5, iPhone S5)
and two non-FIDO compliant devices (Hamster Area Scanner, Validity
Swipe Sensor).   For each of these two device categories,
one device had a swipe sensor and one device had an area sensor.
One attack created a ``fake finger;'' the other produced a latent fingerprint.
The team was able to carry out each attack successfully
on the S5 and on both non-FIDO compliant devices.
Differences between FIDO and non-FIDO compliant devices were not due to the FIDO protocol
but rather to differences in the strengths of the component authenticators.

Subsequently, NSA removed the problem from the INSuRE problem set
because the three student teams had answered all of the questions.

%%%%%%%%%%%%%%%%%%%%%%%%%%%%%%%%%%%%%%%%%%%%%%%%%%%%%%%%%%%%%%%%%%%%%%%%%%%%%%%%%
\medskip \noindent {\it Example 3: Detecting Intrusions on SCADA Systems.}
% Tommy Morris

A two-student team from Mississippi State University studied machine learning techniques for detecting cyber attacks
against industrial control systems~\cite{tom2016}. The team worked from a dataset by Pan~\cite{pan2015} of 
cyber attacks and normal behavior from an electricity transmission system.  
The dataset included alteration and injection attacks against protection relays and {\it Energy Management System (EMS)} software. 
The injection attacks sent illicit network packets to protection relays to cause the relays to operate and open a circuit breaker.  
The alteration attacks used a Man-in-the-Middle to alter voltage and current sensor data sent from 
phasor measurement units to the EMS software.  
The dataset also included instances of single line-to-ground faults at random locations in the simulated transmission system, 
and changes in system load at random times.

First, the team extracted features from the dataset and applied clustering techniques to learn classes of events.
Second, the team built a classifier using the Mamdani fuzzy inference system.
Inputs to the classifier comprised a heterogeneous collection of voltage, current, frequency, Snort log, 
and protection relay log information from one time stamp. 

The team validated their work by comparing results to similar classifiers developed from K-means and Fuzzy C-Means clustering algorithms.  
Their approach outperformed K-means- and Fuzzy C-means-based intrusion detection systems.

%%%%%%%%%%%%%%%%%%%%%%%%%%%%%%%%%%%%%%%%%%%%%%%%%%%%%%%%%%%%%%%%%%%%%%%%%%%%%%%%%
\medskip \noindent {\it Additional Examples.}  %% short
In spring 2014, the INSuRE Project negotiated access to Google Glass by some INSuRE students who analyzed its security.  
A UMBC student team discovered a vulnerability and developed a proof-of-concept demonstration exploit, enabling 
an adversary to activate the eyewear and capture images without the subject's knowledge.

Also in spring 2014, a UMBC PhD student led a group working on a custom project
to perform a security audit of software 
developed for the Random-Sample Election Project.\footnote{\url{http://rsvoting.org/ }}  

A Purdue team investigated information leakage from encrypted output 
from a Cisco ASA 5506 {\it Virtual Private Network (VPN)} appliance. 
The team discovered a vulnerability in the VPN's use of {\it Quick UDP Internet Connections (QUIC)}, 
an experimental transport layer protocol designed in 2012 at Google. 
QUIC exposed the length of short communications, such as passwords, even when encrypted,
as a result of aligning cryptographic block boundaries to contain packet loss. 
When Purdue reported the vulnerability, Google responded that the vulnerability is broader 
than QUIC and also impacts {\it Transport Layer Security (TLS)}. 
If Google’s claim is true, the vulnerability could be used to undermine authentication in TLS 
(for short passwords) and impacts over 96\% of all websites. 
Google and Apple are working with Purdue to create a patch to protect QUIC, TLS, 
and other transport security protocols vulnerable to leaking encrypted password lengths.

Working on the problem of commercial solutions for classified, another Purdue team
explored the impact of known vulnerabilities on layered solutions.
The project analyzed past known vulnerabilities across multiple mechanisms in a layered solution.  
Given known vulnerabilities, as well as patch times generated by statistical models, 
the project simulated the performance of a layered security solution. 
The study also explored if any of the vulnerabilities 
in the individual mechanisms permitted a breach in the security 
to persist beyond the application of a patch, thus allowing 
an attacker to bypass the layered solution even when the vulnerability windows for the layers do not align.  
The team produced software that, 
given one or more layers composing a layered solution, 
identifies windows of opportunity where an attacker could
breach any layer as well as the complete layered solution.
The sponsoring government organization is currently using this tool.

\section{Outcomes and Lessons Learned}
\label{sec:outcomes}

\begin{comment}
In this section we discuss some of the outcomes, benefits, lessons learned, and
open challenges from the INSuRE experience.
\bigskip \noindent {\it Outcomes.}
\bigskip \noindent {\it Benefits.}
\bigskip \noindent {\it Lessons Learned.}
\bigskip \noindent {\it Open Challenges.}
\end{comment}

From fall 2012 through fall 2016, the INSuRE class 
produced 140 project reports on 110 separate problems, 
and taught 356 students (many of whom have been hired by government organizations).

In addition to the works presented at INSuRE-Con, INSuRE projects resulted in 
refereed conference publications (e.g., \cite{alves2016,nair2016,tho2016,alabi2015,cho2015,falk2015}),
refereed posters (e.g., \cite{shehab2014,raut2014}), and
published datasets (e.g., \cite{dark2015c}).

In this section we discuss some of the takeaways (outcomes, benefits, lessons learned, and
open challenges) from the INSuRE experience, 
organized from the separate perspectives of educators and government policy makers.

\subsection{Takeaways for Educators}
We summarize some of the outcomes, lessons learned, and challenges 
from the perspective of educators.

\medskip \noindent {\it Outcomes.}
To improve the course, faculty frequently discussed process and outcomes.
In May 2014, students were asked to submit course feedback via an online survey administered through SurveyMonkey.      
Items included rank-order, Likert scale, and open-ended questions.  
Students rated the course high on a variety of indicators.  
The most highly rated elements included developing expertise in a specific topic in cybersecurity, 
developing qualifications for a job in cybersecurity, and working with a mentor from government or industry.  
Students identified development of research skills in cybersecurity as an important course outcome.  
Results of the survey also showed that students found limitations 
with the electronic communication methods used to interact with other institutions.

In fall 2016, Purdue University conducted a pilot study investigating 
the effect of the INSuRE course on student research self-efficacy. 
Research self-efficacy~\cite{bie1996} is a self judgement of one's ability to perform particular research tasks. 
The study included 17~students (5~undergraduate, 12~graduate)
from eight universities that responded to pre- and post-surveys.
Each student measured research self-efficacy 
using a 100-point Likert scale (0~denotes complete uncertainty; 100~denotes complete certainty). 

Given the small sample size and relative nature of Likert scores, the team analyzed the
data using a nonparametric Wilcoxon Test.
Results show that student research self-efficacy improved:
pre-test \hbox{mean} 73.56, \hbox{median} 76.33, \hbox{interquartile range} [\hbox{65.38--83.54}];
post-test \hbox{mean} 83.27, \hbox{median} 86.83, \hbox{interquartile range} [\hbox{74.54--89.42}].
These gains were statistically significant ($z = -2.58$, $p < 0.01$). 
Cronbach Alpha for each survey was 0.96.

%pre-test $\hbox{mean} = 73.56$, $\hbox{median} = 76.33$, $\hbox{interquartile range} = [\hbox{65.38--83.54}]$;
%post-test $\hbox{mean} = 83.27$, $\hbox{median} = 86.83$, $\hbox{interquartile range} = [\hbox{74.54--89.42}]$.
% table?

Students gained valuable experiences 
carrying out research, 
presenting their work,
writing proposals and reports,
using tools (including for software analysis),
working in groups,
building relationships,
and communicating succinctly and effectively with their TD.
Because the problems touched a broad range of issues, 
students and faculty gained knowledge
outside of their focused areas of expertise.
In addition, the course inspired students to tackle
challenging problems.  For some, it was their first
exposure to research.
The project-based course helped students learn the need to take the initiative 
and lead.

Several INSuRE students continued their studies at the PhD level, citing their INSuRE course
as an important motivating factor.  One university reports that its INSuRE course prompted a faculty
member to include one of the INSuRE problems to his research area.    
This university also reports that the INSuRE course
increased the number of students completing capstone engineering projects in cybersecurity
and motivated more local companies to engage in cybersecurity projects with the university.

The INSuRE course has benefited from a significant number of female students.

Although the INSuRE research experience inspired most students, a few learned
that cybersecurity research is not a path they wish to pursue---a useful discovery for the students.

\medskip \noindent {\it Lessons Learned.}
Many factors contributed to the success of groups.  
To begin, it was helpful to screen students (especially the undergraduates) 
to make sure they are motivated and ready to engage in research.
It was helpful for each team to have a student
leader with strong organizational skills.  
TDs also contributed significantly
through their enthusiasm, availability, and probing questions.

Course alumni and alumnae also contributed to project success.
Some of them enthusiastically functioned 
as course assistants, facilitators, and mentors.

In some terms, instructors required each group 
to provide periodic peer evaluations of a paired group.  
Doing so provides additional feedback to the
evaluated group, and it helps the evaluating group learn the
research process.  While there can be significant value, such
peer evaluations come at a cost of time and effort for
students and faculty, and it can be difficult to coordinate across
diverse university schedules. 

Some schools restricted enrollment to graduate students;
others permitted some undergraduates to participate.  
In mixed classes, more advanced (e.g., PhD) students were
usually expected to carry out greater leadership roles than
were undergraduates.
Many of the instructors found, however, that student performance
typically had more to do with student capability and motivation than
with degree level.  

While most schools offered the INSuRE experience as a dedicated course, others
enrolled students as independent study projects or as part
of an existing capstone course.

% (Univeristy of Houstin).

The faculty found the biannual in-person meetings very useful,
helping participating universities 
to improve the course by applying lessons learned.
They also fostered stronger personal relationships necessary for effective collaboration.  

\medskip \noindent {\it Challenges.}
Challenges included dealing with different time zones
and university schedules (e.g., semester vs. quarter systems).
Also, while useful, the conferencing software yielded video
displays that were limited in comparison to the rich interaction
possible through in-person meetings.   It requires a significant 
involvement by the instructor to stay on top of all projects.

One semester is a short period of time to complete a research project, 
yet one year might be longer than many students are
willing to invest.

At many of the universities, grant support was essential to enable
a faculty member to be allowed to teach a small specialized research course counting
toward his or her official teaching duties.  Teaching the INSuRE class often
meant not teaching some other course, which might have been a larger
required class.  

Some centralized support was essential, to organize the network,
to maintain a repository of project work, and
to manage the collaborative technologies.   Financial resources are needed
for this centralized support, 
for hardware and software, 
for the in-person meetings, 
for support by a teaching assistant if the class is large,
and for instructor time.

\begin{comment}
[true but weak---are we going to do?  how would we do it?]
It would be useful to develop comprehensive evaluation strategies 
to gather evaluative data of the approach and to document project outcomes.
\end{comment}

%\subsection{Takeaways for Researchers}
% Government officials 

\subsection{Takeaways for Government Policy Makers}
At modest investments, the INSuRE Project produced a sizable return, especially in terms of 
recruiting highly-qualified cybersecurity students into the government workforce.  Beyond resumes, 
student experiences with INSuRE demonstrate the ability of students to understand technology,
communicate, and work effectively in teams.  In addition, by funding a research network, 
government can support cybersecurity research without favoritism to particular universities.

The INSuRE course enabled government organizations to stimulate research on projects that they
lacked time to pursue.

% UMBC REDACTED

Aspects of the INSuRE model can be applied to other settings.
Inspired in part by the INSuRE Project, UMBC is pioneering a new
initiative in which UMBC extends SFS awards
to students at nearby Montgomery College and Prince George's Community College, 
who will complete their degrees at UMBC.
While in community college, the scholars will work collaboratively to help solve 
IT security problems for their county government.

Securing a sustainable funding model is a challenge.  One option is a subscription model
in which companies and organizations contribute in return for access to students
and their work on problems.  Another model is a charity model in which sponsors
(e.g., government) fund the program for the national good.
INSuRE welcomes the opportunity to explore future relationships with 
government, industry, foundations, and other groups
to continue the outstanding student work nurtured by INSuRE.

%%%%%%%%%%%%%%%%%%%%%%%%%%%%%%%%%%%%%%%%%%%%%%%%%%%%%%%%%%%%%%%%%%%%%%%%%%%%%%%%%%%%%%%

\section{The Future}
\label{sec:future}

The INSuRE Project has inspired and educated students in cybersecurity,
empowered them to think freely, and engaged them in research problems
related to national security.  It has also strengthened the CAE-R network
and helped government organizations, including by motivating students 
to pursue government service.
The course is being offered again in spring 2017.
INSuRE's continued success will depend on strong external 
support from government, industry and foundations, and internal support from universities.

\section*{Acknowledgments}

We thank Mark Loepker (NSA) for overall government program management and 
the many technical directors who contributed greatly to the success of the INSuRE Project, 
especially Trent Pitsenbarger (NSA).
We also thank the INSuRE instructors and
student leaders.  

This work was supported in part by the National Science Foundation under 
a variety of grants including an Eager grant (1344369), 
supplements to several SFS and other grants
(1241576, 1433753, 1433690, 1241668, 1433795,
1241576, 1241668).
NSA supported INSuRE~2.0 via collaborative CAE-R grants to
DSU (H98230-15-1-0298), 
Stevens (H98230-15-1-0299), and Purdue (H98230-15-1-0300).  
A small Intel grant to Purdue paid for equipment for some student projects.

%\begin{comment}

Additional technical directors include:
Paul Black (NIST),
Joel Coffman (APL),
Nate Evans (ANL),
Aaron Ferber (ORNL),
Jennifer Fowler (ANL),
Andrew Gearhart (APL),
Dickie George (APL),
Mahantesh Halappanavar (PNL),
Al Holt (NSA),
Chris Jenkins (Sandia),
Mark Krisanda (NSA),
Matt Kuczynski (NJOHSP),
Bill LaCholter (APL),
Bill Layton (NSA),
Mark Loepker (NSA)
Joshua Lyle (ANL),
Brian Lyles (ORNL),
Patrick Mackey (PNL),
Mike Moore (ORNL),
Nick Multari (PNL),
Alex Newman (NJOHSP),
Rob Nichols (APL),
Nate Paul (ORNL),
James Pendergass (APL),
Austin Roach (Crane),
Andy Sampson (NSA),
Myron Schlueter (AIS),
Michael Thompson (ANL),
Roland Varriale (ANL),
Andy White (NSA),
Jian Yin (PNL), 
Ed Zieglar (NSA), and
Neal Ziring (NSA).

In addition to the faculty coauthors, INSuRE faculty include:
Matt Bishop (UC Davis),
Glenn Dietrich (San Antonio),
Tom Halverson (DSU),
Michael Ham (DSU),
Doug Jacobson (ISU),
Brandeis Marshall (Purdue),  
Wayne Pauli (DSU),
Joshua Strolchen (DSU), and
Patrick Tague (CMU).

In addition to the student coauthors, INSuRE student leaders include:
Courtney Falk (Purdue),
Filipo Sharevski (DePaul), and
Lauren Stuart (Purdue).

%\end{comment}

%\clearpage
\small
\bibliography{mybib}
\bibliographystyle{plain}

\bigskip \noindent
A shorter version of this paper will be submitted to {\it IEEE Security \& Privacy}.

%\clearpage
%\tableofcontents

\end{document}